\documentclass[sigconf]{acmart}

\acmConference[ICSE 2024]{46th International Conference on Software Engineering}{April 2024}{Lisbon, Portugal}
\usepackage{amsmath}
\usepackage{graphicx}
\usepackage{booktabs}
\usepackage{balance}
\usepackage{xcolor}
\usepackage{xspace}
\usepackage{dsfont}
\usepackage{caption}
\usepackage{subcaption}

\newcommand{\pink}[1]{\textcolor{black}{#1}}

\usepackage{multirow}
\usepackage{array}

\AtBeginDocument{%
  \providecommand\BibTeX{{%
    Bib\TeX}}}

\setcopyright{acmcopyright}
\copyrightyear{2023}
\acmYear{2023}
\acmDOI{XXXXXXX.XXXXXXX}

\acmJournal{JDS}
\acmVolume{37}
\acmNumber{4}
\acmArticle{111}
\acmMonth{8}

\newcommand{\emptystate}{{\sc Empty State}}
\newcommand{\richstate}{{\sc Rich State}}

\newcolumntype{P}[1]{>{\raggedleft\arraybackslash}p{#1}}
\copyrightyear{2024}
\acmYear{2024}
\setcopyright{acmlicensed}\acmConference[ICSE-SEIP '24]{46th International Conference on Software Engineering: Software Engineering in Practice}{April 14--20, 2024}{Lisbon, Portugal}
\acmBooktitle{46th International Conference on Software Engineering: Software Engineering in Practice (ICSE-SEIP '24), April 14--20, 2024, Lisbon, Portugal}
\acmDOI{10.1145/3639477.3639729}
\acmISBN{979-8-4007-0501-4/24/04}

\begin{document}

\title{Enhancing Testing at Meta with Rich-State Simulated Populations}

\author{
Nadia Alshahwan, Arianna Blasi, Kinga Bojarczuk, Andrea Ciancone, Natalija Gucevska, Mark Harman, Simon Schellaert, Inna Harper, Yue Jia, Michał Królikowski, Will Lewis, Dragos Martac, Rubmary Rojas, and Kate Ustiuzhanina, Meta Platforms Inc.
}
\renewcommand{\shortauthors}{Alshahwan et al.}
\acmArticleType{Review}
\acmCodeLink{https://github.com/borisveytsman/acmart}
\acmDataLink{htps://zenodo.org/link}

\begin{abstract}
This paper reports the results of the deployment of Rich-State Simulated Populations at Meta for both automated and manual testing.
\pink{
We use simulated users (aka test users) to mimic user interactions} and acquire state in much the same way that real user accounts acquire state.
For automated testing, we present empirical results from deployment on the Facebook, Messenger, and Instagram apps for iOS and Android Platforms.
These apps consist of tens of millions of lines of code,  communicating with hundreds of millions of lines of backend code, and are  
used  by over 2 billion people every day. 
Our results reveal that rich state increases average code coverage by 38\%, and endpoint coverage by 61\%. 
More importantly, it also yields an average increase of 115\% in the faults found by automated testing.
The rich-state test user populations are also deployed in a (continually evolving) Test Universe; a web-enabled simulation platform for privacy-safe manual testing, which has been used by over 21,000 Meta engineers since its deployment in November 2022.
\end{abstract}
\keywords{Software Testing, Cyber Cyber Digital Twins, Simulation-Based Testing, Machine Learning }

\maketitle

\section{Introduction}

In system level testing, a  test user is typically required.
The test user plays the role of a real user.
For less interactive systems, 
this may have little impact, other than the need for the test user to be logged in.
However, when real users interact with the system, they typically accrue state (e.g., history of purchases).
For such stateful systems, it is additionally necessary for the test users to mimic this behaviour, in order to achieve full test coverage and fault
revelation.

To simulate community interactions, test users also need to interact with {\em one another} through the platform.
This form of testing (with highly interactive test users) is becoming increasingly important as systems, themselves, increasingly become interactive platforms on which communities of users interact~\cite{gray:everything}. 
For example, platforms for online shopping, 
the `gig' economy, and  
for social media and communications,
all involve interaction between users.
On such platforms, users interact with each other through the system, thereby accruing state. 
System-level testing thus requires populations of test users that model such community behaviours.

As the test user modelling of real user behaviours becomes more sophisticated, 
testing enters the realm of Simulation Based Testing (SBT), 
in which the test system becomes a digital twin of the system under test~\cite{jaetal:ease21-keynote}. 
This is characterised by the need to make test users first class citizens; agents with defined and (for testing purposes) controllable  behaviours and state.

Although there have been a great many studies of automated test data generation~\cite{anand2012automated,cadar:three-decades,mh:icst15-keynote}, there has been comparatively little work reporting on the impact of test user {\em state} on fault revelation.
With this paper we seek to draw the attention of the research community to this important problem for real-world system-level testing, 
illustrate problems, and propose some initial solutions for rich state populations, with directions for future work.

Specifically, we report the results of the deployment of rich-State Simulated Populations at Meta for both automated and manual testing.
For automated testing,
we investigate two different SBT test generation techniques with which test users interact to form simulated communities, thereby accruing state.
We report on the application of these two techniques to three popular social media applications: Facebook, Instagram, and Messenger on two platforms, iOS (for Facebook and Messenger) and Android (for all three apps). 
The rich state approach has been deployed in all five of these apps since 2022.

In the most simple mode of test deployment, at the commencement of a test run, the test users have no initial state (denoted \emptystate).
In the \emptystate~ approach, test users thus behave as though they were a community of real users, who had just joined the platform.
In the other mode of deployment  (denoted \richstate), a full simulation of previous test user interactions is performed on Meta's  WW
 platform for SBT~\cite{ahlgren-etal:wes,jaetal:ease21-keynote}.
The WW simulation executes on the real Meta platform, but with test users in place of real uses \cite{jaetal:ease21-keynote}.
WW thereby evolves the test users' state `organically'  through a simulation that models the way in which real users naturally interact with each other on the platform.
Although it uses the real platform for the simulation, the test users are completely isolated from real users \cite{jaetal:ease21-keynote,kmetal:fausta}.


Our results reveal that the WW pre-evolution of test user state is advantageous.
It achieves higher overall coverage and accrues coverage at a faster rate.
This has a knock-on effect on  the  number of faults revealed. 
This finding applies across all apps studied, 
and on both platforms, indicating that these are reasonably consistent and reliable findings.

We also use \richstate~ Populations to generate an evolving simulation  of the Meta platforms, such as Facebook and Instagram.
This platform, known as the `Test Universe', allows Meta engineers to inhabit test user accounts, and control their associated test user personas in a simulated world. 
The world stimulated by the Test Universe includes rich  synthetic  simulated content, and contains {\em only} test user accounts, thereby ensuring that the Test Universe is both privacy safe and that it is securely isolated from production.
Nevertheless, since the test universe executes on the WW platform~\cite{ahlgren-etal:wes}, it is a highly realistic simulation, in which all interactions use the same full stack as production ~\cite{jaetal:ease21-keynote}.

The primary contributions of this paper are twofold: 

\begin{enumerate}
\item An experience report of the deployment of the Test Universe, its uptake by employees and its application at Meta Platforms Inc.

\item 
A report of the improvements in coverage and fault revelation that result from the deployment of rich evolved test user states.
The results show a significant increase in both coverage and fault revelation, with high effect size for all five apps studied. 
Over all apps, the average increase in fault revelation is 115\%.

\end{enumerate}

\begin{figure}[t]
    \centering
    \includegraphics[width=\linewidth]{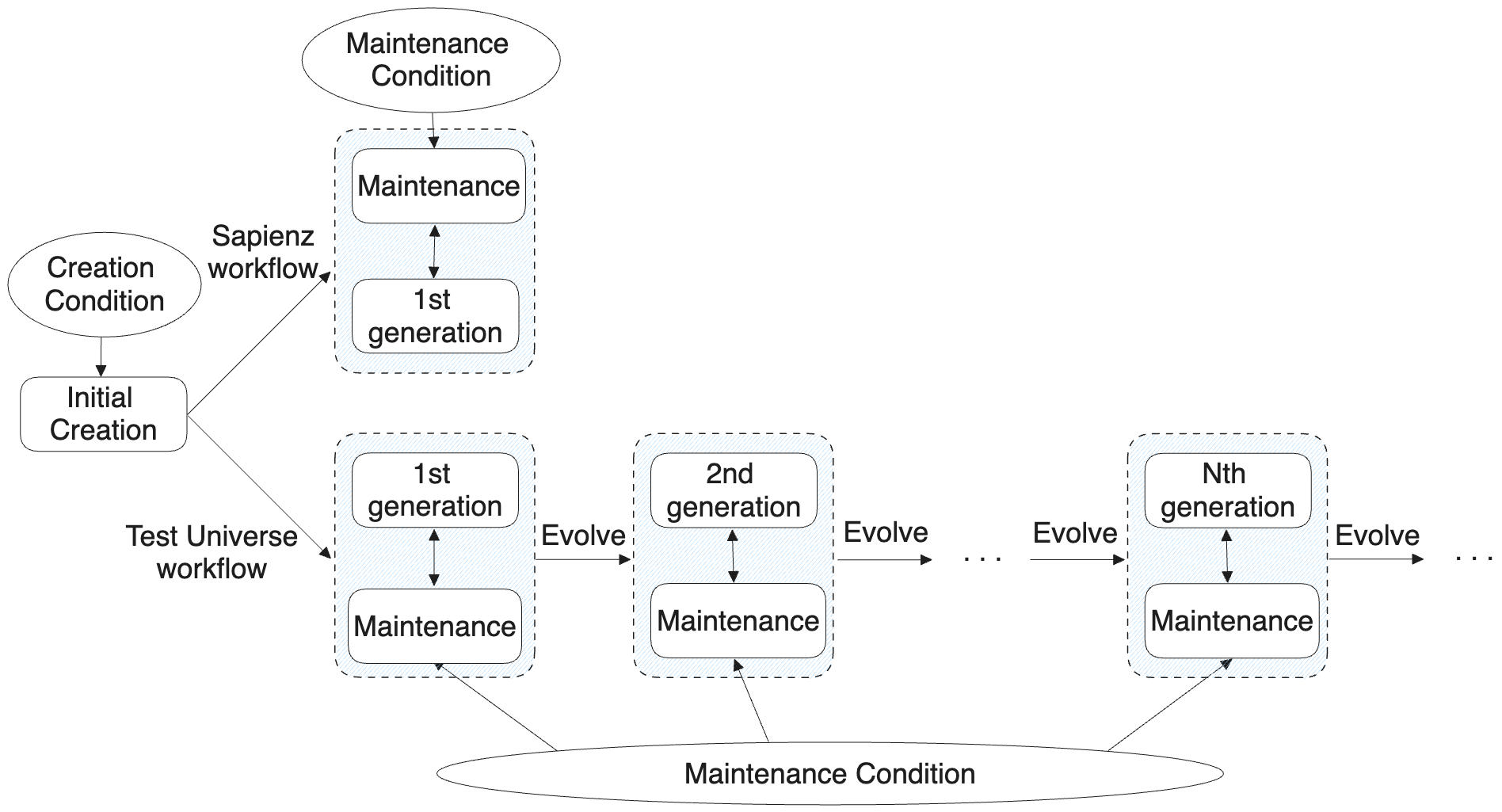}
    \caption{The Populations Manager workflow. This is how \richstate~populations are created and evolved. For Sapienz, we only evolve once on initial creation, and use the 1st generation only. For the Test Universe, we evolve once per day, so we are currently on the Nth Population where N is the number of days since the population has been created.}
    \label{fig:pop_manager_workflow}
\end{figure}

\begin{figure}[t]
    \centering
    \includegraphics[width=\linewidth]{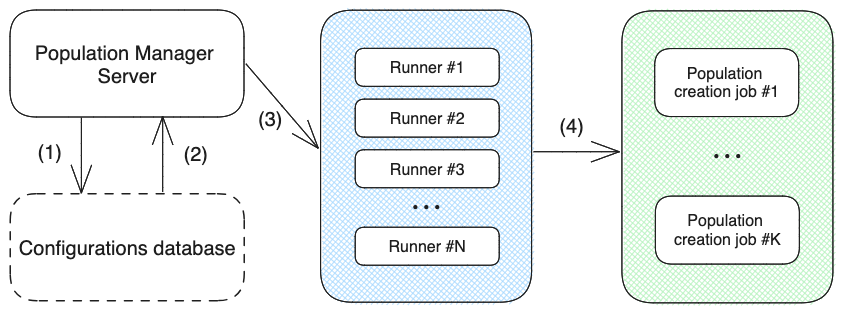}
    \caption{Populations Manager architecture. The server queries the configuration database (1) and receives a list of current populations and their settings (2). For each population, the server maintains a runner (3), that is responsible for monitoring and scheduling new jobs to create additional test users (4).}
    \label{fig:population_manager_architecture}
\vspace{-5mm}
\end{figure}

\begin{figure}[t]
\begin{tabular}{||c r r||} 
 \hline
 Name & \# populations & \# test users\\ [0.5ex] 
 \hline\hline
 Test Universe & 1 & 82,177 \\
 \hline
Facebook & 11 &  55,260 \\
 \hline
Messenger	& 4	 & 16,400 \\
 \hline
Instagram	& 3	& 15,000 \\[1ex] 
 \hline
\end{tabular}
\caption{Sizes of test user populations deployed at Meta in the Populations Manager framework.}
\label{fig:population_sizes}
\vspace{-1mm}
\end{figure}

\section{Populations of test users approach}
\label{sec:populations_manager}

We create synthetic \textit{populations} of test users using the WW simulation platform~\cite{ahlgren-etal:wes}. 
Through interactions with the app -- amongst others: posting, commenting or sending messages -- these test users accumulate state information that is unique to each of them. 
The content created by these test users depends on their specific settings, which we call their test user \textit{personas}.

In order to efficiently control and scale the populations, we implement the Populations Manager: a system responsible for scheduling and monitoring their creation, as well as \textit{maintaining} the populations, i.e., removing test users from the populations and scheduling new creations when \textit{creation conditions} are met.
The Populations Manager supports both automated  and manual testing.
We depict these two workflows in Fig.~\ref{fig:pop_manager_workflow}.
The upper workflow, labelled `Sapienz workflow', depicts the fully automated testing  case.
The lower workflow, labelled `Test Universe workflow', depicts the evolutionary workflow used to support the Test Universe.

Consider the upper workflow.
We use the Sapienz automated test generation system  \cite{mao:sapienz:16,mhetal:ssbse18-keynote} to automatically explore  apps using a population generated in a single generation of the overall evolution process.
This population is updated to refresh and retire test users, ensuring that they remain suitable for automated testing, according to a set of well-defined maintenance conditions.
In the Sapienz workflow, we generate the content once for every active test user. 
Section \ref{sapienz_section} describes how we use Sapienz to automatically explore Meta apps using the population generated by this workflow.

The lower workflow, labelled `Test Universe workflow', is more elaborate, because it has to cater for continual evolution over multiple generations of an interacting population of test users.
This workflow essentially simulates the real Meta platforms, such as Instagram and Facebook.
However, the same overall Populations Manager is used to maintain each generation. 

The principal difference between the Sapienz workflow and the Test Universe workflow is that, in the Test Universe workflow, between each generation, the population {\em evolves}. 
That is, test users interact with one another, for example liking each other's posts, and sending each other messages. 
This interaction creates a test user community, the `Test Universe'.

In the Test Universe, test users play the role of bots that automatically and autonomously interact 
through the evolutionary process managed  by the Populations Manager.
As the test users (bots) interact with each other in the Test Universe,
they accrue additional
test user state.
In this way, their autonomous evolution ensures that their state is also continually updated (evolving), 
simulating the way in which real user state continually evolves through interaction in real user communities.
Section~\ref{test_universe} describes the Test Universe deployment in more detail.

The Populations Manager also orchestrates the maintenance of the populations it creates.
The maintenance conditions vary from use case to use case.
To illustrate, consider the automated generation of tests using Sapienz.
The test users' states can change during the process of automated testing (because testing may cause test users to interact).
This is unhelpful for replication and testability; such interactions can lead to test flakiness. 
Furthermore, certain forms of automated testing are designed to reveal bugs, rather than to faithfully replicate normal user behaviour. 
As a result,
some of the test user states might diverge from being a realistic representation of content on the platform, potentially negatively influencing the test coverage e.g. all new messages get opened and there's no more notifications to read. 
To alleviate this issue, we set a limit on the number of times a particular test user can be used in automated tests. 
After hitting this limit, the test user is deactivated and not used in any future tests.
This limit forms one of the maintenance conditions for the Sapienz workflow, and thereby ensures that test users remain fit for purpose.

\subsection{Test Users}
Test users are first class citizens.
They execute all of their actions using the same full stack of backend systems used by real users, and they observe their world through the same APIs that ultimately feed information to real users on the real platforms. 
In this way, we ensure the realism of the simulation; a founding principle of web enabled simulation~\cite{ahlgren-etal:wes} which makes the simulation essentially a Cyber Cyber Digital Twin \cite{kbetal:esem21-keynote,jaetal:ease21-keynote} of the Meta platforms Instagram, Messenger and Facebook.

Each test user persona specifies the frequency with which each feature (such as posting content, and responding to posts) will be used by the associated test user. 
With this high level of control over the amount and type of content that is included in the final state, 
we can create different populations that target specific features of the tested app. 
For example, when the tester is  mostly interested in features related to Facebook's Marketplace, we can create a population that will be mostly focused on items' listings, as well as various selling and buying activities, making them prominent in the resulting population state.

\begin{figure}
    \includegraphics[width=0.8\linewidth]{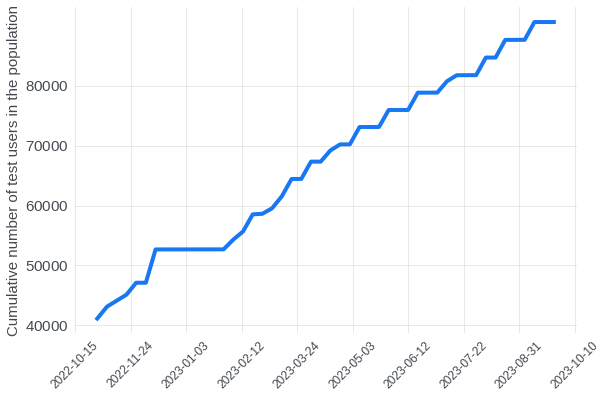}
    \caption{Total number of test users in the Test Universe between November 2022 and August 2023.}
    \label{fig:test_users_in_pop}
    \vspace{-5mm}
\end{figure}

\begin{figure}
    \includegraphics[width=0.8\linewidth]{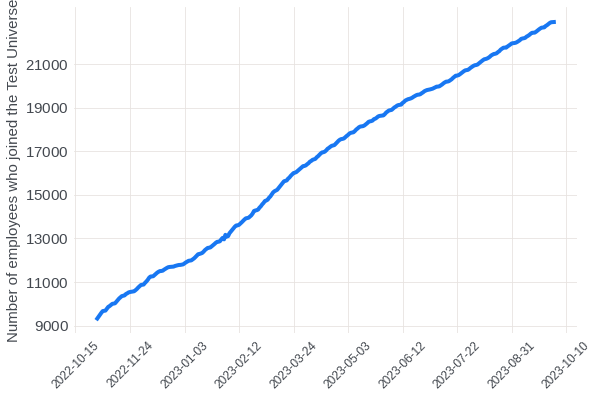}
    \caption{Cumulative number of Meta employees who claimed a test user from the Test Universe between November 2022 and August 2023. As can be seen, there has been a steady rise in adoption over the period of deployment, to the point where over 21,000 employees have claimed a test user.}
    \label{fig:employees_num}
    \vspace{-5mm}

\end{figure}

\subsection{Determining population dynamics}
Population level dynamics are determined by a population  configuration, which specifies, not only its basic properties, such as the size (number of test users), but also the properties of their test user state. 
For the latter, we specify a distribution of different test user personas over the whole population. 
This combination of population configuration, and the distribution of test user personas, uniquely defines the behaviour of the population.
To illustrate, consider a population of test users in which  10\% have their personas set-up to frequently create postings on Marketplace, while the remaining 90\% are more focused on other aspects of the platform. 
This population would be useful for testing Marketplace buy-sell scenarios, 
where we ensure that we represent the Marketplace sellers in the 10\% and the remaining users are just ordinary users who could buy and/or do other things on the platform.
Allowing multiple population configurations for the same application enables us to create isolated environments in which completely different types of experiments can be conducted.

\subsection{Architecture and scale}

We present the architecture of the Populations Manager in Figure~\ref{fig:population_manager_architecture}. 
The Populations Manager runs continuously, polling for new configurations and any changes to the configurations for the existing populations. 
This makes the process of onboarding a new type of population quick and friction-less, as the engineers need only describe the specific test user settings. 

At the time of writing, the Populations Manager maintains 18 different populations (excluding the Test Universe, described in detail in Section \ref{test_universe}) across Facebook, Messenger, and Instagram, with more than 86,000 test users overall. 
Figure~\ref{fig:population_sizes} describes the specific sizes of the populations for each of the apps.

  

\section{Test universe deployment} \label{test_universe}

The Test Universe is designed to support cases where an engineer wants to manually verify the behaviour of the implemented feature in a controlled simulated  environment.
It provides realism and privacy safety, because of its implementation on top of Web Enabled Simulation \cite{ahlgren-etal:wes}.
The Test Universe has  grown steadily and consistently since its deployment, and is now used by over 21,000 employees at Meta.
The growth of the number of test users in this population is depicted in Figure \ref{fig:test_users_in_pop}. 
The growth correlates to the number of claimed test users (see Figure \ref{fig:employees_num}). 
There is a short period of plateau around December 2022. 
This is due to the Christmas holiday period when a lot of  employees were less active due to annual leave.

The Test Universe is primarily designed to support manual testing.
Employees adopt the persona of a test user in the test universe in order to explore the behaviour of existing and new features.
This allows employees to perform system-level testing on the real platform, while entirely isolated from production users in a safe sandbox-style environment.
Despite it being primarily designed for manual testing, the Test Universe does, nevertheless, overcome some of the hurdles that limit the efficiency and effectiveness of manual testing at the system level.

One of these (frequently encountered) hurdles that concerns the need to set up test data objects.
This is typically achieved using data builder APIs.
These APIs provide facilities to construct objects that represent content, such as posts, reels (short form video), and stories.
Additionally, data builders are required to update test user state to reflect properties of test users, such as posts the test user has liked, the test user's friends in the friendship network, and groups that the test user has joined.
While these APIs hide implementation details, a considerable amount of human-written code is required to set up the test user state for a specific test purpose.

As a result, it can take a long time to write tests, making the manual testing process, tiresome, error-prone, and inefficient.
To alleviate this issue, the Test Universe provides engineers with
a set of test users in which state is already present.
The engineer does not have  to explicitly set out, neither in detail nor in code, the exact content and interaction history of a test user.
Rather, using the Test Universe, 
the test engineer can simply set up broad parameters that define the kinds of content and interactions pertinent to the test user. 
This is done through the persona.

The Test Universe  can be thought of as a kind of `second life' version of the platforms, Instagram, Messenger and Facebook, available to all Meta engineers for manual testing. 
Engineers, and other Meta employees, can claim a test user within this network and use them for testing purposes. Claiming means the employee becomes the sole owner of the test user.
In this way, the test engineer sacrifices a small degree of control over the exact detail of interaction history and content, 
for a great deal of reduced effort in test construction.
The steady, consistent and large-scale uptake of the Test Universe revealed in Figure~\ref{fig:employees_num} indicates that the Test Universe strikes the right balance
between detailed test user content/interaction control and ease of test process.

The test user is a kind of `alter ego' for the engineer. 
The engineer takes on the test user and can log in and interact with content and other test users in the test universe, in exactly the same way as they would do using the regular platform.
When an employee claims a test user in the Test Universe, they acquire complete control over settings related to that particular test user: this not only includes the ability to configure persona settings, but also the ability to influence which features will appear in the friend network. 
Additionally, whenever an employee claims a test user, the Populations Manager  automatically connects them with their teammates' test users (linking their test users as friends), 
thereby imbuing the newly-claimed test user with a pre-existing (and natural) network of friends. 
This also allows for easy sharing of test resources (e.g. custom tailored groups with specific content) in the population.

When an employee claims a test user, we call this their `Primary Test User'. 
Employees may also want to control other test users, which they simply want to act as bots that interact with their Primary Test User. 
We therefore maintain three pools of test users:
\begin{enumerate}
\item {\bf Claimed}: Each claimed test user is assigned to a specific employee as the employee's Primary Test User. 
This test user becomes the employee's alter ego in the test universe, and also their way of interacting with their colleagues' Primary Test Users, and with unclaimable bots.  
\item {\bf Unclaimed}: each unclaimed test user continues to evolve autonomously to ensure it will have state when it becomes claimed by an employee as their Primary Test User.
\item {\bf (Unclaimable) Bots}: bots autonomously evolve and interact with all other test users. They cannot be claimed as  Primary Test Users, but they can be claimed as secondary test users, the sole role of which is to interact with Primary Test Users e.g. send messages to the user.
Bots play a role a little bit like non-player characters in interactive game worlds, except that they are, to some degree, {\em controllable} by employees in their interactions with Primary Test Users.

\end{enumerate}

As the maintenance condition of the Test Universe, the Populations Manager strives to maintain a fixed ratio of unclaimed-to-claimed test users, in order to make sure there are sufficient unclaimed test users to not limit the ability to control the friend network content. 
As employees cannot change persona settings for test users claimed by other employees, it is necessary to have a sufficiently large pool of unclaimed test users
and unclaimable bots. 
This is also the responsibility of the Populations Manager.

\subsection{Evolving population}
The Test Universe is a large population of test users with an evolving state that is  updated each day.
The Populations Manager schedules jobs that automatically generate new content for each of the test users in the population according to their specified personas. 

While the daily evolution of the state within the Test Universe is important to ensure that there is enough content for manual testing, 
it is also crucial because  certain features expire after a set amount of time, such as Facebook Stories. 
Combined with persona preferences, this gives Meta employees a rich pool of test objects that are ready to be used on demand.

\subsection{Adoption Process}
The Test Universe was initially released to selected small groups of engineers. 
This allowed us to quickly gather feedback through unstructured interviews with engineers and iterate on the commonly requested features to prepare for the release to all Meta employees.
Once this initial set of features was established, we gradually rolled out and scaled the Test Universe, incrementally on-boarding employees and providing additional features, as needed.

A range of features were introduced based on the continuous feedback from employees gathered in feedback forms as well as 1:1 user studies. 
These features were launched in order to increase further adoption. 
We believe all these features have collectively helped to increase the number of onboarded employees, but for the sake of brevity, we illustrate with two examples, here

\noindent 
{\bf Example 1}:
Feedback revealed how important it was that the content itself should be pleasant to work with.
We found that a form of `test fatigue' set in when much of the textual content, and/or other aspects of content such as video and images, was generated as random synthetic content.
Although the synthetic content was generated to be realistic, and typical of content that might otherwise be generated on the real platform, this was insufficient to avoid fatigue.
Rather, employees using the Test Universe wanted to see content that was also meaningful for {\em them}, much as regular users of the platform might want to do.
We found that this tended to reduce fatigue and increase engagement.
Therefore, in addition to controlling the type of features created by their test user, 
Meta employees are also able to set specific subjects -- which we name test user \textit{interests} -- which will then be used to populate the content, such as relevant text, pictures, and videos.

Employees can include specific items of content, if they choose, but it can be time-consuming to specify content at this level of detail.
We found that specifying content at the higher level of `topics of interest'  proved to be sufficient to generate content that was meaningful to employees.
It yielded content that employees found to be pleasing to engage with, and relevant to the test user's owner, while simple and high-level to specify, thereby being relatively friction-free as a deployment vehicle for employee-relevant content.

\noindent 
{\bf Example 2}:
To decrease the friction of logging into the Test Universe and increase its adoption, we introduced a \textit{profile switcher}, which allows employees to switch between their personal account and their Primary Test User account with a single click from the Facebook UI.

Figure~\ref{fig:employees_num} shows the degree of uptake of test users by Meta employees after the launch. 
As can be seen, deployment has led to a steady linear growth in the size of the Test Universe, and the number of employees using it for their manual testing activities.
At the time of writing (August 2023) there are over 80,000 test users evolving in the Test Universe each day, with approximately 21,000 employees and other Meta employees using or having used a Primary Test User.



\section{Sapienz Rich State Deployment } \label{sapienz_section}
Sapienz ~\cite{mao:sapienz:16,mhetal:ssbse18-keynote} is an unsupervised testing platform that provides autonomous end-to-end testing for Meta's family of apps.
It was first deployed in 2017, as a search based automated test generation platform to test all Meta products \cite{mhetal:ssbse18-keynote}.
The initial deployment targeted the Android platform, for which Sapienz automatically designs test cases, executes them, and reports failures in 
Meta's continuous integration environment.
In 2019, Sapienz was extended to also generate test cases for the iOS platform.
Since these initial deployments, both the range of failures targeted, and the algorithms used to target them, have been considerably extended and adapted.
Initially, to circumvent the Oracle Problem \cite{ebetal:oracle}, Sapienz focussed purely on crashes, but has since been extended to tackle memory issues and performance-related regressions.

The algorithms used to uncover faults have also been extended to include many other exploratory strategies, including reinforcement learning.
By using reinforcement learning, Sapienz explores the app's features without the need for human input.
Sapienz Exploration Mode focuses on automatically exploring the app's features to maximize code coverage and identify faults. 
We use Sapienz Exploration Mode to continuously test the master builds of Meta's apps and create crash-fixing tasks for the engineering teams.


Since 2022, Meta has extended the Sapienz deployment to use the \richstate~ test user population to further augment its pre-production fault-revealing potential.
The \richstate~ allows the Sapienz algorithm to take advantage of realistic user content and connections to speed up the testing process with more realistic test scenarios. 
For example, Sapienz will spend more time testing Facebook Groups features if the test user belongs to many groups, while it may tend to test more messaging features when a test user has many friends.


\begin{figure*}[t]
\centering
  \begin{tabular}{ |p{2.5cm}||P{0.8cm}|P{0.7cm}||P{0.8cm}|P{0.7cm}|||P{0.8cm}|P{0.7cm}||P{0.8cm}|P{0.7cm}|||P{0.8cm}|P{0.8cm}|P{1.0cm}|P{1.0cm}|}
    \hline
    \multirow{4}{*}{App Name} &
      \multicolumn{4}{c|||}{\small{Endpoints}} &
      \multicolumn{4}{c|||}{PLs} &
      \multicolumn{2}{c|}{Increase} &  \multicolumn{2}{c|}{Wilcoxon P-value} \\
     & \multicolumn{2}{c||}{Unique}  & \multicolumn{2}{c|||}{Total} & \multicolumn{2}{c||}{Unique}  & \multicolumn{2}{c|||}{Total} & \multirow{3}{0.8cm}{\centering End points} & \multirow{3}{0.8cm}{\centering PLs}  & \multirow{3}{1.0cm}{\centering End points} & \multirow{3}{1.0cm}{\centering PLs} \\
    & \emptystate & \richstate & \emptystate & \richstate & \emptystate & \richstate & \emptystate & \richstate & & & & \\
    \hline
    Facebook Android & 28 & 239 & 451 & 662 & 48 & 435 & 1270 & 1657 & 47\% &  30\%  & 0.000 & 0.000\\
    Facebook iOS & 18 & 198 & 244 & 424 & 26 & 369 & 697 & 1040 & 74\% &  49\%  & 0.000 & 0.000\\
    Messenger Android & 6 & 40 & 124 & 158 & 15 & 101 & 584 & 670 & 27\% &  15\% & 0.016 & 0.016\\
    Messenger iOS & 4 & 32 & 104 & 132 & 10 & 87 & 398 & 475 & 27\% &  19\% & 0.014 & 0.014\\
    Instagram Android & 45 & 183 & 204 & 342 & 62 & 321 & 477 & 736 & 68\% &  54\% & 0.009 & 0.009\\
    \hline
    All (sum) & 101 & 692	& 1127	& 1718& 	161	& 1313	& 3426& 4578 & 61\% &  38\% & &\\

    \hline
  \end{tabular}
\caption{Increase in coverage achieved by \richstate~ over \emptystate~ in terms of (i) Endpoint coverage and (ii) PL coverage over each of the different apps. 
As can be seen, both forms of coverage are notably improved using the \richstate~approach.
This data is depicted as a  Venn diagram in Figure~\ref{fig:coverage_venn}.The P values are for a paired Wilcoxon test wrt the Null Hypothesis that the results obtained with \richstate~ are not significantly different to those with \emptystate. } 
\label{fig:table_coverage_increase}
\end{figure*}

\begin{figure*}[t]
\centering
  \begin{tabular}{ |p{2.5cm}||P{0.7cm}|P{0.7cm}||P{0.7cm}|P{0.7cm}|||P{0.7cm}|P{0.7cm}||P{0.8cm}|P{0.9cm}|||P{1.2cm}|P{1.2cm}|P{1.4cm}|  }
    \hline
    \multirow{4}{*}{App Name} &
      \multicolumn{4}{c|||}{Crashes found in Single Build} &
      \multicolumn{4}{c|||}{Crashes found in Multiple Builds} &
      \multicolumn{2}{c|}{Increase} & \multirow{4}{1.3cm}{\centering Wilcoxon P-value for Single Build}\\
     & \multicolumn{2}{c||}{Unique}  & \multicolumn{2}{c|||}{Total} & \multicolumn{2}{c||}{Unique}  & \multicolumn{2}{c|||}{Total} & \multirow{3}{1.2cm}{\centering Single Build} & \multirow{3}{1.2cm}{\centering Multiple Builds} & \\
    & \emptystate & \richstate & \emptystate & \richstate & \emptystate & \richstate & \emptystate & \richstate & & & \\
    \hline
    Facebook Android & 8 & 67 & 21 & 80 & 0 & 266 & 120 & 386 & 281\% &  222\% & 0.000 \\
    Facebook iOS & 2 & 5 & 4 & 7 & 0 & 7 & 24 & 31 & 75\% &  29\% & 0.000\\
    Messenger Android & 3 & 11 & 9 & 17 & 0 & 13 & 50 & 63 & 89\% &  26\% & 0.022\\
    Messenger iOS & 4 & 8 & 10 & 14 & 0 & 9 & 43 & 52 & 40\% &  21\% & 0.026\\
    Instagram Android & 1 & 3 & 3 & 5 & 0 & 5 & 23 & 28 & 67\% &  22\% & 0.019\\
    \hline
    All (sum) & 18 &94	&47	&123	&0	&300	&260	&560 & 161\% &  115\% & \\
     \hline
  \end{tabular}
\caption{Increase in \richstate~ over \emptystate~ in Failures found over (i) 1 build and (ii) 10 builds from August 2023. 
As can be seen, the \richstate~ approach significantly improves fault revelation.
This data is depicted as a  Venn diagram in Figure~\ref{fig:crashes}.The P values are for a paired Wilcoxon test wrt the Null Hypothesis that the results obtained with \richstate~ are not significantly different to those with \emptystate.} 
\label{fig:table_crashes_increase}
\end{figure*}

\begin{figure}\captionsetup[subfigure]{font=footnotesize}
     \begin{subfigure}[b]{0.30\textwidth}
         \centering
         \includegraphics[width=\textwidth]{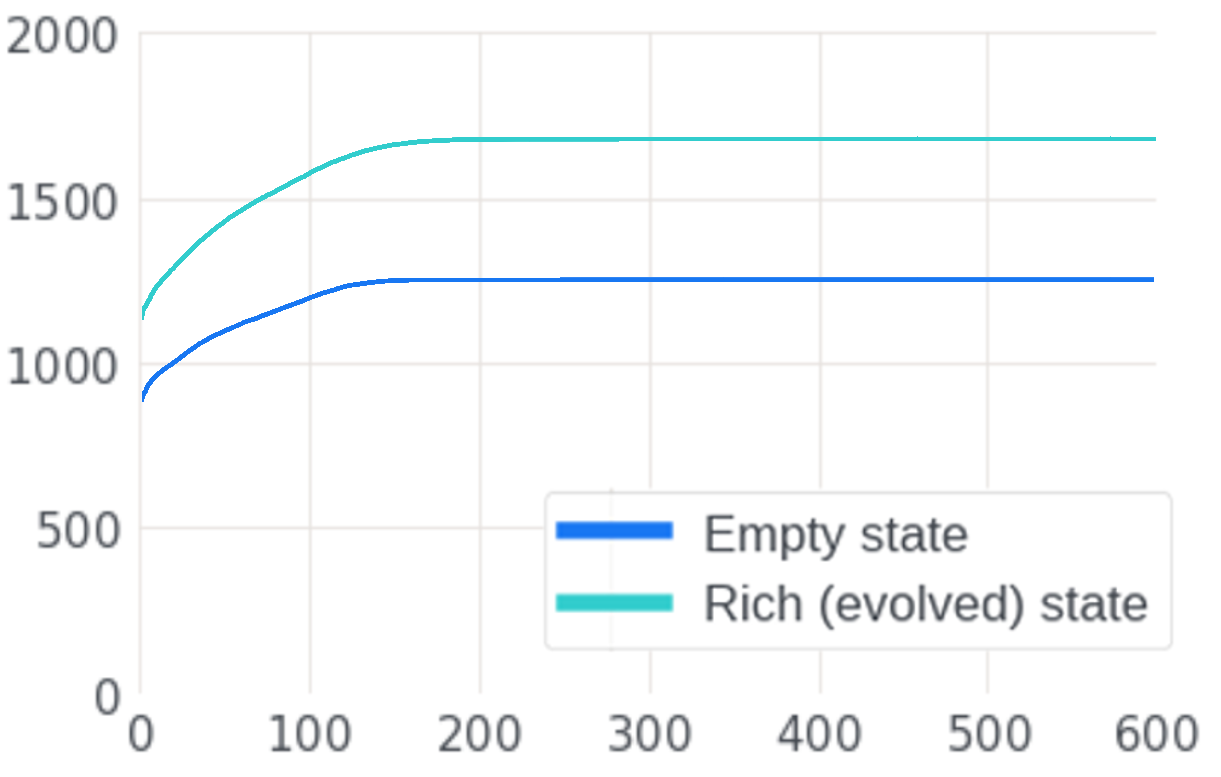}
         \subcaption{Facebook App on Android.}
         \label{fig:qpls_time_fba}
    \vspace*{-1mm}
     \end{subfigure}

    \begin{subfigure}[b]{0.3\textwidth}
         \centering
         \includegraphics[width=\textwidth]{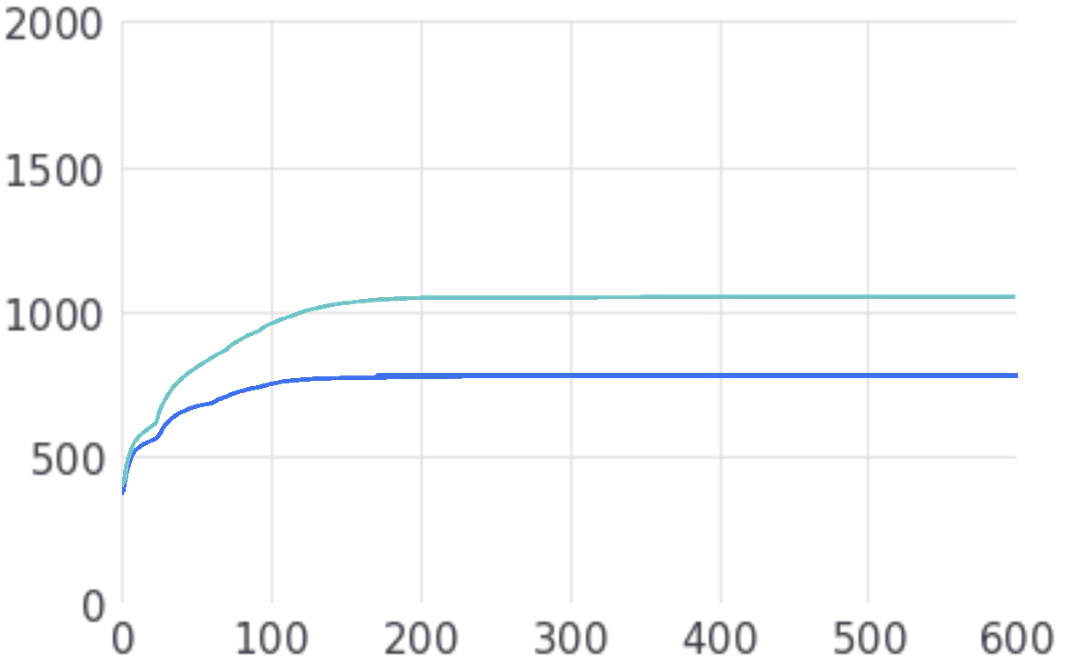}
         \caption{Facebook App on iOS.}
         \label{fig:qpls_time_fbios}
             \vspace*{-1mm}
     \end{subfigure}
         \begin{subfigure}[b]{0.3\textwidth}
         \centering
         \includegraphics[width=\textwidth]{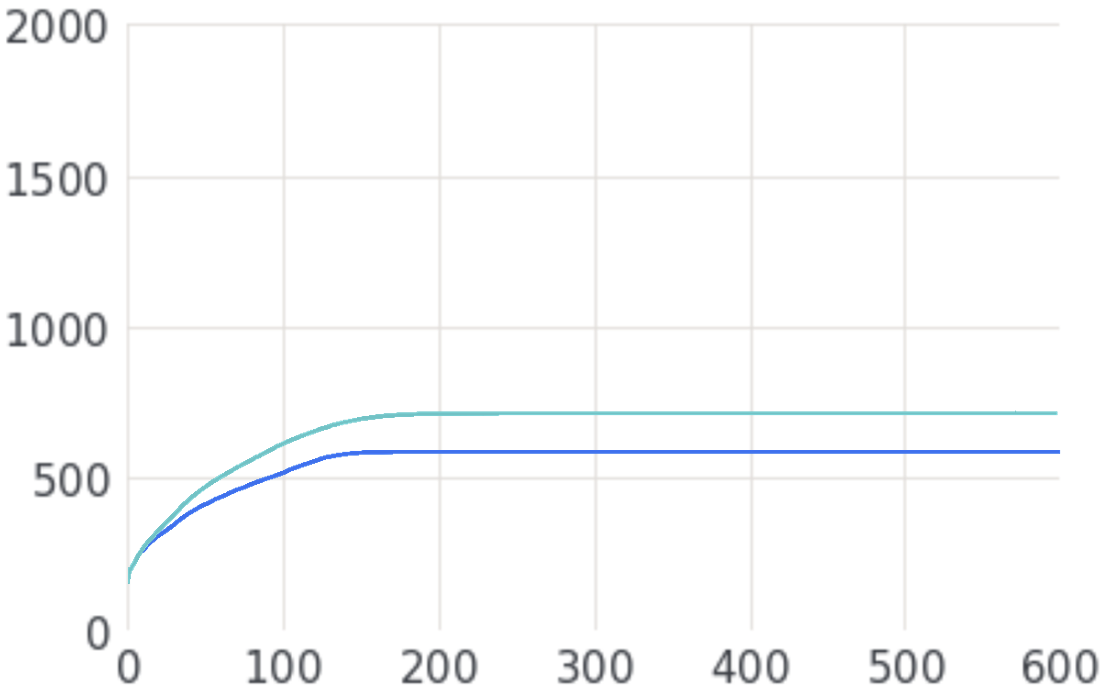}
         \caption{Messenger App on Android.}
         \label{fig:qpls_time_mess_a}
         \vspace*{-1mm}
     \end{subfigure}
    \begin{subfigure}[b]{0.3\textwidth}
         \centering
         \includegraphics[width=\textwidth]{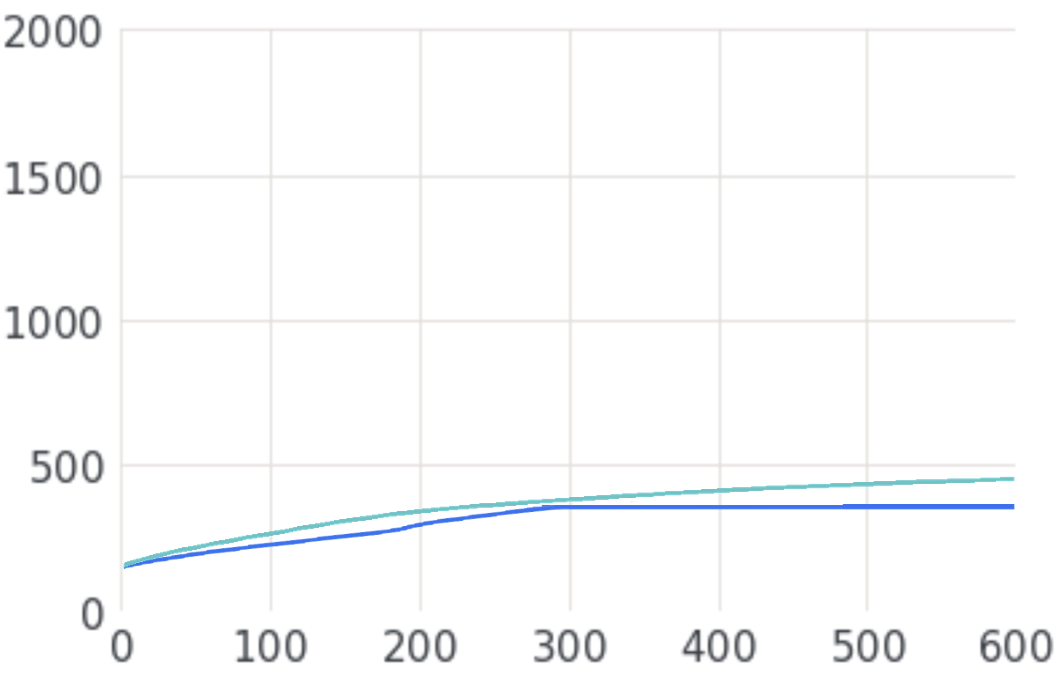}
         \caption{Messenger App on iOS.}
         \label{fig:qpls_time_mess_ios}
         \vspace*{-1mm}
     \end{subfigure}
        \begin{subfigure}[b]{0.3\textwidth}
         \centering
         \includegraphics[width=\textwidth]{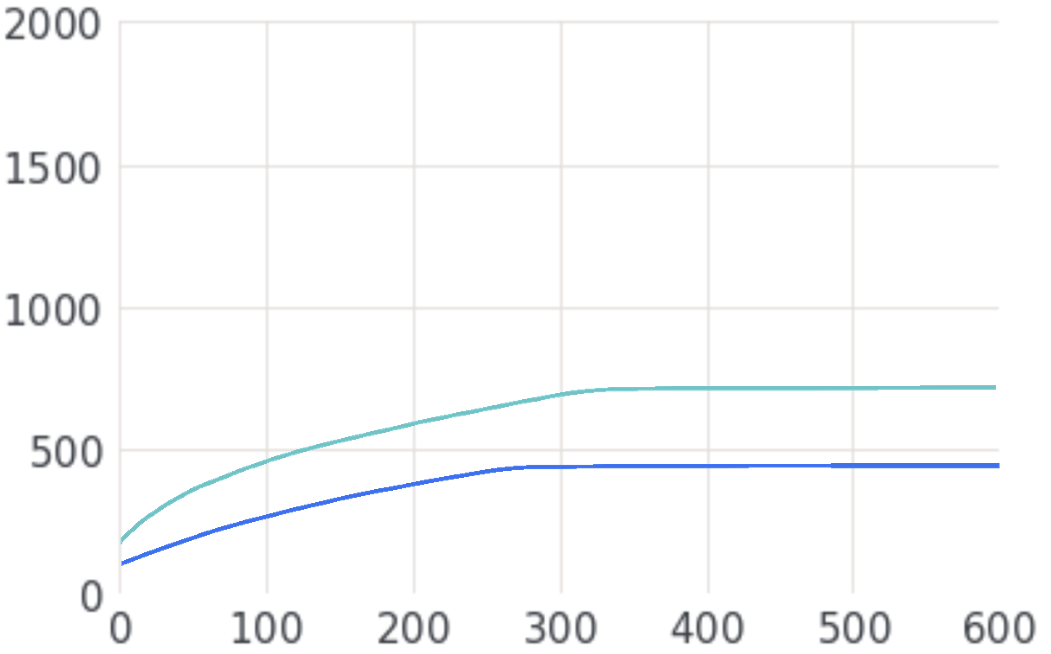}
         \caption{Instagram App on Android.}
         \label{fig:qpls_time_ig}
         \vspace*{-1mm}
     \end{subfigure}
     
\caption{Code level coverage growth over the duration of the test generation process, measured by PL (performance logger) events (results are similar for endpoints). 
Test automation is orchestrated by test exploration with Sapienz. 
The upper line in each sub-figure depicts coverage obtained using {\richstate}, while the lower is obtained using {\emptystate}. 
As can be seen, the coverage achieved grows faster, and further, using the {\richstate}~approach, giving evidence for the benefit of the \richstate.}
\label{fig:qpls_time}
\end{figure}

\section{Evaluation}
\label{sec:evaluation}
In this section, we present empirical results to evaluate the advantage conferred by the use of the \richstate~ on the deployment of Sapienz. 
To do this, we compare against a shadow deployment with the \emptystate~ in an otherwise identical deployment context.
We  ran experiments on  Alpha builds constructed  between July and August 2023. 
An Alpha build is an internal release of a version of the app containing multiple changes from different development teams.
All Alpha builds are tested by Sapienz, and by other testing infrastructure before being allowed forward into the Beta release process. Once the previous app version is cut for the Beta Branch release, the major app version for the new Alpha release branch is incremented at approximately one-week intervals. To ensure that the builds we are testing include a significant number of different changes\footnote{A change is landed into the code base approximately every few minutes \cite{mhpoh:scam18-keynote}. }, we select the initial Alpha build from the major app version update. The release process ultimately leads to submission to the corresponding App Store (Android or iOS).

By running in shadow mode deployment with the \emptystate~on all Alpha builds,  
we are able to give a fair comparison, as if the \emptystate~ approach  had been deployed into the release process.
This is what we mean by `shadow deployment'.
We thus directly compare Sapienz runs on  \emptystate~Population vs \richstate~ Population in a like-for-like setting,  on a realistic set of  builds of the app.

We call a single Sapienz exploration on a specific app a `Sapienz run'. 
A Sapienz \textit{deployment} dictates several configuration parameters about runs. 
In particular, we define how many runs will be assigned to a specific Meta app and on which build specifically, and what kind of test user population Sapienz will use. 
After Sapienz logs into the app with a test user account, it enters the following overall exploration loop:
\begin{enumerate}
\item Extract the UI layout to discover the UI views.
\item Find all possible actions that can be executed on the available views (tapping, scrolling, etc.).
\item Select an action to execute.
\item Check whether the run should complete (i.e., whether the test process has used up the pre-defined test budget).
\end{enumerate}

To evaluate the algorithm, we measure  app coverage using both PLs and endpoints. 
A full instrumentation (able to collect fine-grained coverage information), simply does not scale to the size of apps deployed by Meta.
Therefore, we use PLs (Performance Loggers).

PLs are lightweight probes inserted into the app code, 
that collect information on key features covered in the code while not unduly impacting app performance or size.
\textit{PLs} are the preferred way, at Meta, to log duration and outcome of specific \textit{events}, and to assess whether an interaction is successful from the user's perspective.
They also provide a convenient and lightweight way to measure code coverage.
\textit{Endpoints} are the entry points into a system or application, typically a mobile app screen. 
These correspond well to interactions between the client and server, and therefore, also represent a form of coverage.
In this section, we report both PLs and endpoints to measure the coverage achieved at the system level by a test framework, such as Sapienz.

Of course, coverage is necessary, but not sufficient, for good testing.
While high coverage can give us confidence in the signal provided by the test process, 
the effectiveness of a testing strategy must also be measured by its ability to reveal faults.
At the system level, Sapienz identifies failures, and uses a triage mechanism to trace these back to faults, 
reporting the corresponding fault to engineers as a signal in the Meta Continuous Integration environment \cite{mhetal:ssbse18-keynote}.
The mapping between faults and failures can be a subtle one, and there are occasionally duplicates.
Nevertheless, our experience is that the triage process ensures a reasonably close to one-to-one mapping between failures and the faults reported to engineers as test signal.
Counting unique system level failures found, such as app crashes triaged to unique causes,  
thus provides an effective proxy for the number of unique faults found by the test process.

\subsection{Coverage}
\label{sec:coverage}
We seek to answer the following research questions about coverage achieved by the \richstate: \textbf{RQ1:} What is the improvement in coverage in \richstate~ Population over the \emptystate? We divided this question into 2 sub-questions:

\begin{itemize}
  \item \textbf{RQ1.1} How does the PL coverage achieved by \richstate~compare to the PL coverage achieved by \emptystate?
 \item \textbf{RQ1.2:} How does the endpoint coverage achieved by \richstate~compare to the endpoint coverage achieved by \emptystate?
  \item \textbf{RQ1.3:} What is the rate at which the PL coverage grows over the duration of the test process with \richstate~ and \emptystate~approaches?
\end{itemize}

\begin{figure}[t]
    \begin{subfigure}[b]{0.45\linewidth}
         \centering
         \includegraphics[width=0.7\textwidth]{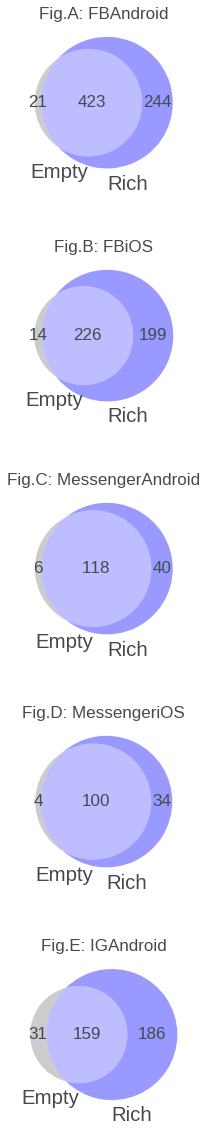}
    \caption{Unique Endpoints covered.}
    \label{fig:final_unique_endpoints}
     \end{subfigure}
      \begin{subfigure}[b]{0.45\linewidth}
         \centering
         \includegraphics[width=0.7\textwidth]{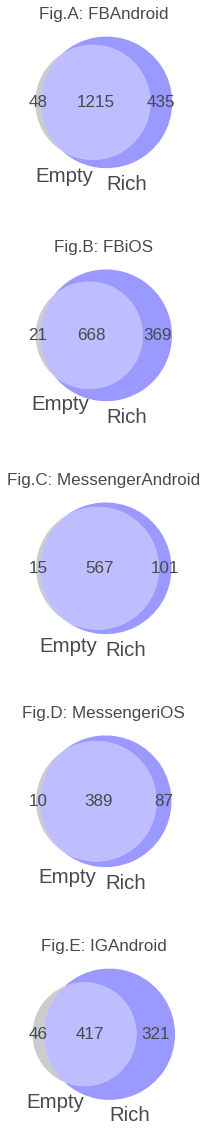}
    \caption{Unique PLs covered.}
         \label{fig:final_unique_qpls}
     \end{subfigure}
    \caption{Unique coverage metrics across \emptystate~ and \richstate~ within Sapienz Automated Testing Runs. \richstate~ clearly provides a bigger coverage, however it does not cover everything \emptystate~ did.}
    \label{fig:coverage_venn}
\vspace{-5mm}
\end{figure}

To answer these coverage questions, we calculated the average of coverage over 2000 runs per 10 different  Alpha builds. 
The table in Figure~\ref{fig:table_coverage_increase} presents the overall results for coverage achieved by the two approaches.
Figures \ref{fig:final_unique_endpoints} and  \ref{fig:final_unique_qpls} provide a more visual representation of the unique coverage 
achieved by each approach, and the intersection, in terms of endpoints and PLs covered by both approaches.

The overall improvement in coverage is 38\% on average for PLs, to 61\% on average for endpoints.
Furthermore, for each app and for each way of measuring coverage, the coverage achieved by the \richstate~ approach outperforms that achieved by the \emptystate~approach.
The smallest coverage improvement is a 15\% PL improvement in coverage for Messenger on Android, while the largest improvement is the 74\% endpoint coverage improvement for Facebook on iOS.

In general, it is notable that the coverage improvements tend to be stronger for Facebook and Instagram compared to Messenger. 
We believe that this is due to the richness of features available in Facebook and Instagram, compared to Messenger.
That is, Messenger is essentially a messaging product in which users can send messages to one another.
By comparison, Instagram and Facebook include messaging, but also many other social media interactions.
We must naturally be careful to avoid  overgeneralising based on only three apps.
Nevertheless, these observations may provide some additional evidence that the rich state is increasingly important for apps that have higher levels of 
content-based user interaction.

We might have thought that coverage due to the richer state would {\em subsume} that achievable in an empty state. 
However, the results do not show this.
It is, therefore, interesting to dive deeper into these results for the \emptystate~approach.
Upon more detailed manual inspection, we found two primary reasons why the \emptystate~approach may cover endpoints and PLs not covered by the \richstate~approach:

\begin{enumerate}
\item {\bf Features for onboarding new users}: There are some code paths that are specific to new users' journeys, such as  onboarding flows for new users. 
\item {\bf Increased probability of hitting deeply nested features}: Certain pages are  deeply nested within the app structure. 
    For example, as the pages that allow changes to settings.
    Since these are so deeply nested, the code and endpoints corresponding to each individual setting may have a low probability of being hit by exploratory testing.
    However, in the empty state, the algorithm has fewer choices, since there are fewer features available e.g. no posts, no messages that are clickable etc.
    For each setting available, in the empty state, the exploration therefore has a slightly higher overall chance of being hit during exploratory testing.
\end{enumerate}

Figure \ref{fig:qpls_time} presents the results for RQ1.3. 
Each graph plots the unique cumulative coverage achieved.
Each data point plotted is the average cumulative coverage over 10 runs.
We use averages in order to cater for non-determinism present in an individual run.
All ten graphs plotted in this figure show a typical logarithmic growth in test coverage.
This is a typical (and expected)  pattern that is witnessed in many testing scenarios because, 
as cumulative coverage is achieved, 
there are fewer remaining available uncovered items \cite{anandetal:orchestrated,mhpm:issta07,zeller:beautiful}.

More importantly, the graphs show that, when testing with  \richstate~ Populations, 
test exploration is able to achieve faster coverage growth than when testing with  \emptystate.
This is important, because all the automated test generation algorithms deployed with Sapienz are essentially `anytime' algorithms \cite{zilberstein:using}; they accrue value over time, and can be terminated at any point. 
This is typical for all testing processes, so many test generation algorithms share this `anytime' characteristic.
In terms of software testing, the test generation algorithm typically terminates when the test budget is exceeded. 
However, any faults  found during the process of testing are reported immediately. 
Therefore, a faster coverage growth rate will typically translate into earlier signal to the engineer from the test process.


To further test the observations we make about the superiority of the \richstate~ Populations, we perform a paired Wilcoxon inferential statistical test.
In each case, the observations are paired by the Alpha build from which they are constructed: 
one using the \richstate~ Population, and one using the \emptystate~ Population.
As recommended by standard tutorial advice on inferential statistical analysis of search based (and other nondeterministic) algorithms \cite{arcuri:practical,mhetal:sbse-tutorial},
we use the Wilcoxon test, because this is a non-parametric test, and we cannot be sure that the data is normally distributed.
In this way, we obtain a P value for paired observations for each app; 
five P values for the PL-based coverage measurement and five for the endpoint-based coverage. 

These P-values are shown in Table~\ref{fig:table_coverage_increase}.
As can be seen, in all cases, the P value is lower than 0.05, indicating that we can  reject, at 95\% level, the Null Hypothesis that the coverage observed from the \richstate~ Populations is no better than the coverage achieved by the \emptystate~ Populations.
In this case, the effect size is so strong, that it is revealed by only ten data points, 
and since the Null Hypothesis is rejected in every pairwise case (i.e., for each app), 
there is no risk of Type II error, even with this relatively low sample size.
Furthermore, since in every case, the \richstate~ produces higher coverage than the \emptystate~ Populations, 
the non-parametric Vargha-Delaney test \cite{neumann:transformed}  for effect size is 1.0 in every case, 
suggesting the highest possible effect size.

In conclusion, the overall answer to RQ1 for coverage, is that 
automated testing using \richstate~ Populations  is significantly superior to the baseline using \emptystate~ Populations.
With the richer test user state, testing is able to achieve higher overall coverage (of both code markers (PLs), and of endpoints) and it does this  on all apps studied and on both Android and iOS platforms and, in all cases, it also achieves a faster growth rate of coverage.

\subsection{Failures}
We seek to answer the following research question in terms of failures: \textbf{RQ2:} What is the improvement in terms of failures found in \richstate~ Populations vs \emptystate~ Populations. 
We focus specifically on crashes, 
since these tend to be the most unequivocal of all system failures.
We have divided the answer into 2 sub-questions:
\begin{itemize}
 \item \textbf{RQ2.1} Are there any type of crashes that can only be detected in \richstate~ and not in \emptystate ? 
 \item \textbf{RQ2.2} When testing on a single build, are there any crashes that are detected faster by one method than by the other? 
\end{itemize}

\begin{figure}[t]
    \begin{subfigure}[b]{0.49\linewidth}
         \centering
         \includegraphics[width=0.65\textwidth]{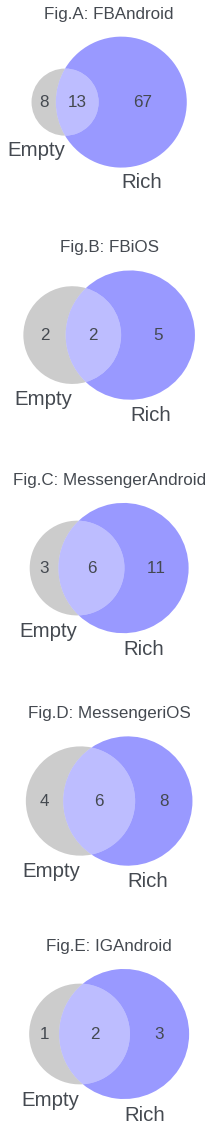}
    \caption{Unique failures found in one particular build.}
    \label{fig:crashes_one_build}
     \end{subfigure}
      \begin{subfigure}[b]{0.49\linewidth}
         \centering
         \includegraphics[width=0.63\textwidth]{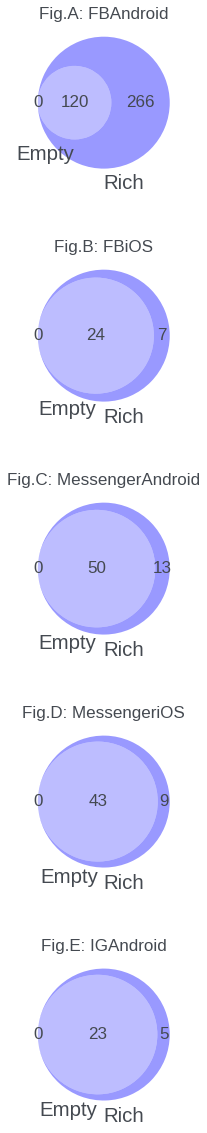}
         \caption{Unique crashes found across multiple builds}
         \label{fig:crashes_multiple_builds}
     \end{subfigure}
    \caption{Number of unique failures found across \emptystate~ and \richstate~ within Sapienz Automated Testing Runs. Overall, more crashes were found in \richstate. Although, sometimes \emptystate~ found them faster.}
    \label{fig:crashes}
\vspace{-2mm}
\end{figure}

We measure the total number of failures detected over 10  Alpha builds between July and August 2023. 
The table in Figure~\ref{fig:table_crashes_increase} presents the overall results for the crashes found in an (arbitrary) single build of the app under test (averaged over ten runs), and the crashes cumulatively found over a consecutive series of ten  builds.
Figures~\ref{fig:crashes_one_build} and~\ref{fig:crashes_multiple_builds} depict the results as Venn Diagrams.

Over multiple builds, there are many unique crashes that can only be found in \richstate~ Populations, 
while there are {\em no} crashes that can {\em only} be found in \emptystate~ Populations. 
This suggests that, in terms of fault detection, \richstate~ Populations are much more valuable since they discover more failures (and thereby more faults caused by these failures). 

To answer RQ2.2 consider Figures~\ref{fig:crashes_one_build} and \ref{fig:table_crashes_increase}. 
The results presented in these figures show the average number of crashes detected in an arbitrary single  
build over 2000 tests. 
The results indicate that there is a subset of crashes that is discovered faster by \emptystate. 
However, RQ2.1 suggests that these crashes can also be discovered by \richstate~ Populations, over time. 
Overall, the results indicate  that the \richstate~approach  is much better at finding failures than the \emptystate~approach,
while the empty state may add value when only a `single shot', non-continuous, testing approach is possible 
(such as when using a  smoke test).

As with coverage, we perform a paired Wilcoxon  test on the data points collected from single builds for  the numbers of crashes found.
The P-values are shown in Table~\ref{fig:table_crashes_increase}.
As can be seen, the \richstate~approach is significantly better than the \emptystate~approach for  all  five apps studied, although the \emptystate~approach can find {\em some} crashes in each single build that are not found by the \richstate~approach.



\subsection{Limitations and Threats To Validity}
The results  presented are relatively unequivocal in their demonstration of the value of rich test user state but, as always, care is required in generalising from these findings.
All Sapienz deployments with {\emptystate}~ Populations and {\richstate~} Populations were run using the same automatic testing algorithm. 
Results may not generalise to other automated testing algorithms.
The applications on which we evaluated are Social Media applications with a substantial amount of code handling  user state, 
such as online interactions, browsing user-generated content, and notifications. 
Therefore, having a {\richstate~} Population 
might be expected to have a substantial benefit during testing. 
Other apps, especially less complex apps with little user state, will be likely to witness {\em some} degree of improvement when state is accrued through user interaction, but the results may differ.

There are other potential threats to the validity of the results.
In order to provide baseline data, we  compare results from production with those from a shadow baseline deployment.
This gives us experimental control, while  ensuring that the shadow deployment closely mirrors the production deployment.
We carefully replicated the exact steps in the shadow deployment that are taken in the production deployment, and the same infrastructure was used for both.
This minimises threats to validity, but cannot remove them entirely.

Since automated testing algorithms are generally  `anytime algorithms' \cite{zilberstein:using},
an important experimental choice is the cut-off; the point at which we stop the test process and report coverage and fault revelation
data achieved up to that point.
We chose to conduct experiments consisting of 2000 runs.
We found this to be sufficient to produce robust experimental results. 
We based this choice on our observation that Sapienz  cumulative coverage acquisition has fully stabilized after 1500 runs. 
Indeed, this claim is borne out by the coverage growth graphs presented in Section~\ref{sec:coverage}.
As a result, we believe that there is a very low probability that additional 
coverage would be achieved after more than 2000 runs, thereby justifying our choice of this is a cut-off.

A further source of potential threats to validity comes from the nondeterministic nature of the underlying test exploration process.
To cater for this, 
we have repeated the experiments over 10  Alpha builds between July and August 2023 and 
used standard inferential statistical testing  techniques to assess significance and effect size. 
We believe this gives sufficiently robust results.

\section{Future work and open challenges}
The problem of catering for test user  state, especially in  automated test generation, is a relatively less well studied problem.
As  demonstrated by the empirical results presented  in Section~\ref{sec:evaluation}, better handling of test users in general, (and test user state in particular) can have significant impact on coverage and fault revelation for highly interactive software systems and platforms.
There are three avenues for future work on which we would be delighted to collaborate with the wider research community:

\noindent
{\bf New Test User State}: 
We have used the simulation-based approach to generate test user state, but there may be other approaches that can generate rich novel content. 
One promising area might be the use of Large Language Models \cite{mhetal:LLM-survey}, which are naturally generative, 
and could be used to generate both content and user interactions equally realistically, but without the full simulation overhead.


\noindent
{\bf
Failure Reproducibility}: A failure may be the result, not only of the state of an individual test user, but of an entire set of test users. 
The existing literature on test failure reproducibility is promising \cite{bell:chronicler,bianchi:reproducing, jin:bugredux}, but it does not yet fully tackle this problem.
For instance, simply cloning the test user and repeating the sequence of actions that caused the crash may  be insufficient, because the crash may depend on the states of their friends. 
To better reproduce such crashes, future systems need to reconstruct a sub-test-user population that replicates the social network states from a previous point where the crash occurred.
We need more work on algorithms for reproducibility in the presence of communities of interacting test users.
Such work will help to identify algorithms that identify minimal needed conditions for reproducibility, 
thereby increasing actionability, and reducing time for debugging and test resources required for reproduction.

\noindent
{\bf
Simulation Algorithm}: Automated test generation algorithms have typically  not considered the state of interacting test user populations as an input to the algorithm, nor as an optimisation goal \cite{anandetal:orchestrated,mcminn:survey}.
However, this may mean that such  algorithms cannot fully explore features that are not reflected by test users' profiles, nor those that require specific sequences of user interactions to have taken place. 
We need more work on algorithms that are aware of test user state and interaction history, and can tune automated testing behaviour to optimise for it.
We expect that such work could have significant impact on testing highly interactive systems.

\section{Related work}
Automated software test generation has a long history, dating back to pioneering work on search based and symbolic execution based test generation, for which there are many comprehensive surveys~\cite{mh:icst15-keynote,cadar:three-decades,mcminn:survey}.
As the field evolved and matured, the research community's ambition widened from  simply covering control flow in small unit size portions of code~\cite{boyer:select,king:thesis,mhetal:tse-flag,fraser:mutation-driven}, to larger system-level test generation~\cite{gross:false,mao:sapienz:16}. 
At the system level, it becomes important to consider, not only the code directly under test, but the backend systems with which  this code communicates. 
Among the back end systems of interest for test generation, stateful systems (such as databases) present particular challenges~\cite{taneja:moda,abdul:automated}.

Automated testing has been well studied in the literature, while platforms to support manual testing more effectively have been less well studied.
Both automated and manual testing need realistic test data content~\cite{anandetal:orchestrated}, and this is all the more important when the system under test involves complex user interactions involving this content.
For example, the data used to populate the database for a test case have to be realistic in order that any failures detected by the test will prove to be {\em actionable} by engineers.
An unrealistic test will not be actionable because it will not denote a failure that an engineer can imagine occurring in production.
These kinds of `unrealistic' test failures risk becoming regarded as false positives~\cite{mhetal:ssbse18-keynote}, 
and consequently being de-prioritised, thereby wasting engineer and test effort.
Even for those few engineers brave enough to tackle complex failures with unrealistic test data, the problem of debugging test failures becomes far more time-consuming, and thereby expensive, because of the lack of use case familiarity and domain context.

Automatically generating {\em realistic} test cases also presents its own challenges: it moves the problem from merely generating test data that covers hard-to-cover paths (a challenge in itself), to that of generating coverage-triggering test data such that it is additionally representative of production data  ~\cite{draheim:realistic,mustafa:ssbse12,alonso:arte}.

Test data realism is especially challenging where the test data of interest concerns user data, for which  privacy considerations mean that these data cannot come from production observations.
This combination of the need to test at the system level, to generate realistic data for backend databases, and to do so in a fully privacy-safe manner, naturally leads us to a Simulation Based Testing approach~\cite{kmetal:fausta,jaetal:ease-keynote,jaetal:mia,tuli:simulation}.

In this paper, we use simulation of user communities, and their interaction to elevate coverage for testing of platform based systems in which users can interact and thereby accrue state.
Simulations have been used to analyse a wide range of human activities, including studies of economics \cite{terzi:sim-survey-economics},
climate change \cite{johnson:stochastic}, 
traffic safety \cite{alsultan:sim-survey-road}, 
and pandemic dynamics \cite{adam:covid_sim-survey}.
Simulation has also been used in the context of social media.
For example,
Serrano et al.~\cite{Serrano2015} survey 18 publications of  rumour spreading simulation on Twitter, while
Luna and Pennock~\cite{luna2018social} discussed  applications in emergency management and
Padilla et al.~\cite{padilla2014leveraging}    analyse agent-based simulations.

The focus of the present paper is on simulation as a way to {\em test} platform-based applications such as social media.
In this context, the simulation can be thought of as a digital twin ~\cite{jaetal:ease21-keynote}.
One potential challenge in using simulation in this way, is the lack of reproducibility~\cite{weiler2019towards,EASE2021Keynote} in the consequent impact on the Oracle Problem of Software Testing ~\cite{ebetal:oracle}.
Not only is the correct behaviour of the system unknown, it is inherently {\em unknowable} \cite{jaetal:mia}.
We sidestep this oracle issue, by focusing purely on the Implicit Oracle  ~\cite{ebetal:oracle}; behaviour that is known to be incorrect in any context, such as app not responding, crashes, and  exceptions (such as out of memory errors).

\section{Conclusion}

This paper introduced the problem of test user state, with a particular 
emphasis on the rich states required to adequately cover and reveal faults in large-scale complex systems involving interactions between multiple users.
The paper provides empirical evidence for the importance of adequately modelling rich test user state in order to achieve adequate system  coverage and fault revelation.
The paper presented results on three popular apps (Facebook, Messenger and Instagram) over two popular platforms (iOS and Android).
The results show that testing with this enriched state consistently and convincingly outperforms the unenriched baseline in terms of both coverage achieved and faults revealed.
The paper also reports on Meta's deployment and uptake of the rich state Test Universe over the period from November 2022 to August 2023.

\bibliographystyle{ACM-Reference-Format}
\bibliography{references,simulation_references}

\end{document}